\documentclass{aa}
\usepackage{txfonts}
\usepackage{natbib}
\usepackage{graphicx}
\bibpunct{(}{)}{;}{a}{}{,}

\begin{document}
\title{Multiple eruptions from magnetic flux emergence}

\author{D. MacTaggart \and A.W. Hood}

\institute{School of Mathematics and Statistics, University of St Andrews, North Haugh, St Andrews, Fife, KY16 9SS, Scotland}

\abstract {} {In this paper we study the effects of a toroidal magnetic flux tube emerging into a magnetized corona, with an emphasis on large-scale eruptions.  The orientation of the fields is such that the two flux systems are almost antiparallel when they meet.}{We follow the dynamic evolution of the system by solving the 3D MHD equations using a Lagrangian remap scheme.}{Multiple eruptions are found to occur.  The physics of the trigger mechanisms are discussed and related to well-known eruption models.} {}


\titlerunning{Multiple eruptions from magnetic flux emergence}
\authorrunning{D. MacTaggart et al.}
\maketitle

\section{Introduction}
The trigger mechanism for large-scale eruptions, such as coronal mass ejections (CMEs), is one of the main concerns of theoretical solar physics.  There are many different approaches to modelling solar eruptions, ranging from 2D analytical models to 3D numerical simulations.  A review of such models can be found in \citet{forbes00} and further references are given in \citet{dmac09a}.  One highly influential model for solar eruptions is the \emph{breakout model} \citep{antiochos99}.  This theory uses, as an initial condition, a multipolar configuration, which consists of four distinct flux systems separated by a null point.  This configuration is stressed by an \emph{imposed} shearing at the lower boundary (normally taken to be the photosphere).  Reconnection across the current sheet at a stressed null changes the field geometry and weakens the tension of overlying field lines, converting them into nonrestraining field lines of neighbouring flux systems.  The continuation of this process can eventually lead to expulsion of a flux rope.  For more details of the breakout model, the reader is directed to \citet{antiochos99}, \citet{devore08} and \citet{lynch08}.

Recently, \citet{devore08} and \citet{lynch08} have simulated magnetic breakout in 3D.  The first of these studies finds homologous confined eruptions.  The second study follows the topological evolution of a fast breakout CME.  \citet{soenen09} simulate homologous CMEs in the solar wind in an axisymmetric 2.5D configuration.

The breakout studies mentioned above have been very useful in investigating the physical mechanisms of solar eruptions.  One point which they all have in common, however, is that the initial equilibrium is stressed by artificially imposed shearing motions.  

Other models are also used to study eruptions in the solar atmosphere.  For example, \citet{mackay06} model the large-scale coronal field as a series of non-linear force force-free equilibria.  Flux ropes form but not all of them settle into equilibrium.  Those that diverge from equilibrium are ejected.

For studying dynamic evolution in the atmosphere, another model is \emph{flux emergence}.  This considers the early evolution of active regions, which, of course, are the sources of large-scale solar eruptions.  The standard practice in dynamic flux emergence experiments is to have a stratified solar atmosphere including the top of the solar interior.  A flux tube is placed in the solar interior and the system is left to evolve by itself, with no imposed flows.  A review of such models can be found in \citet{archontis08rev}.  Some studies include a magnetized corona in their model.  \citet{archontis05} and \citet{galsgaard07} study the effects of a cylindrical flux tube emerging into a horizontal coronal field.  They consider different orientations for the magnetic fields and discuss the reconnection and high-speed jets that occur.  \citet{maclean09} consider the same experiment but investigate the magnetic topology of the system.  In relation to solar eruptions, \citet{manchester04} report on the eruption of a flux rope that forms during the emergence of a cylindrical flux tube into an non-magnetized corona.  \citet{archontis08} study the emergence of a cylindrical flux tube into a magnetized corona and find that with a favourable orientation, a CME-like eruption is possible.

In the present paper, we shall consider a similar setup to \citet{archontis08} but use a different model for the magnetic field.  This is a toroidal loop, placed in the solar interior, rather than a cylindrical one \citep{hood09}.  A comparison of these two models is discussed in \citet{dmac09b}.  This paper will show that multiple large-scale eruptions are possible from the same emerging region and will discuss the physics of the trigger mechanisms.

The outline of the paper is as follows: $\S 2$ will describe the model setup and the initial conditions.  In $\S 3$ we shall discuss the results of the eruption experiment and link the processes involved with eruption models.   $\S 4$ will summarize the results.

\section{Model setup} 
To study the evolution of the system numerically, we use a Lagrangian remap scheme \citep{arber01}.  All variables are non-dimensionalized with photospheric values: pressure, $p_{\mathrm{ph}} = 1.4\times 10^{4}\,\, \mbox{Pa}$; density, $\rho_{\mathrm{ph}} = 3\times 10^{-4}\,\,\mbox{kg}\,\,\mbox{m}^{-3}$; temperature, $T_{\mathrm{ph}} = 5.6\times 10^3\,\, \mbox{K}$ and scale height $H_{\mathrm{ph}} = 170 \,\,\mbox{km}$.  The other units used in the simulations are: time, $t_{\mathrm{ph}} = 25\,\,\mbox{s}$; speed, $u_{\mathrm{ph}} = (p_{\mathrm{ph}}/\rho_{\mathrm{ph}})^{1/2} = 6.8\,\,\mbox{km}\,\,{s}^{-1}$ and magnetic field $ B_{\mathrm{ph}} = 1.3\times 10^3\,\,\mbox{G}$.  The evolution of the system is governed by the following resistive and compressive 3D magnetohydrodynamic (MHD) dimensionless equations
\[
   \frac{\partial \rho}{\partial t} + \nabla\cdot (\rho\mathbf{u}) = 0,
  \]
\[
 \rho\left(\frac{\partial\mathbf{u}}{\partial t} + (\mathbf{u}\cdot\nabla)\mathbf{u}\right) = -\nabla p + (\nabla\times\mathbf{B})\times\mathbf{B} + \rho\mathbf{g},
\]
\[
 \frac{\partial\mathbf{B}}{\partial t} = \nabla\times(\mathbf{u}\times\mathbf{B}) + \eta\nabla^2\mathbf{B},
\]
\[
 \rho\left(\frac{\partial\varepsilon}{\partial t} + (\mathbf{u}\cdot\nabla)\mathbf{\varepsilon}\right)  = -p\nabla\cdot\mathbf{u} + \eta j^2,
\]
\[
\nabla\cdot\mathbf{B} = 0,
\]
with specific energy density $\varepsilon = {p}/{((\gamma-1)\rho)}$.  The basic variables are the density $\rho$, the pressure $p$, the magnetic field vector $\mathbf{B}$ and the velocity vector $\mathbf{u}$. $j$ is the magnitude of current density and $\mathbf{g}$ is gravity (uniform in the $z$-direction). $\gamma$ is the ratio of specific heats and is taken as 5/3.  $\eta$ is the resistivity which is taken to be uniform with a value of 0.001. The code accurately resolves shocks by using a combination of artificial viscosity and Van Leer flux limiters.  In such regions, heating is added to the energy equation.

The equations are solved in a Cartesian computational box of (non-dimensional) sizes, $[-50,50] \times [-50,50] \times [-20,85]$ in the $x,y$ and $z$ directions, respectively.  The boundary conditions are closed on the top and base of the box and periodic on the sides.  The background stratification is similar to that in \citet{dmac09b}.  The non-magnetic stratification includes a solar interior that is marginally stable to convection $(-20 < z < 0)$, a photosphere/chromosphere $(0 < z < 10)$, a transition region $(10 < z <20)$ and a corona $(20 < z < 85)$.  A horizontal magnetic field of the form
\[
 \mathbf{B} = B_c(z)(1,0,0),
\]
is included in the corona.  $B_c(z)$ is a hyperbolic tangent profile, so that the field is uniform in the corona and rapidly declines to zero at the base of the transition region. The strength of the coronal field is taken to be 0.01 ($\approx$ 13G).  The orientation of the field is chosen so that it is almost antiparallel to the field of the emerging flux tube when they meet.  The initial toroidal tube, that is placed in the solar interior, has the form
\begin{eqnarray*}
B_x &=& B_\theta (r)\frac{s-s_0}{r}, \\
B_y &=& -B_\phi (r)\frac{z-z_0}{s} - B_\theta(r)\frac{x}{r}\frac{y}{s}, \\
B_z &=& B_\phi(r)\frac{y}{s} - B_\theta (r)\frac{x}{r}\frac{z-z_0}{s},
\end{eqnarray*}
with
\[
 r^2 = x^2 + (s-s_0)^2,\, s-s_0 = r\cos\theta,\, x = r\sin\theta,
\]
and
\[
 B_\phi = B_0e^{-r^2/r_0^2}, \quad B_\theta = \alpha r B_\phi = \alpha B_0 r e^{-r^2/r_0^2}.
\]
$s_0$ is the major axis of the tube and $r_0$ is the minor axis. $z_0$ is the base of the computational box.  $\alpha$ is the initial twist and $B_0$ is the initial axial field strength. Varying these parameters can have a profound effect on the behaviour of toroidal emergence \citep{dmac09b}. For this study, however, we shall only consider the values of $B_0$ = 5 and $\alpha$ = 0.4. In this study we take $s_0 = 15$, $r_0 = 2.5$ and $z_0 = −25$. To initiate the experiments, the entire tube is made buoyant. i.e. a density deficit relative to the background density is introduced \citep{hood09,dmac09b}.

\section{Results}
\begin{figure}
 \resizebox{\hsize}{!}{\includegraphics{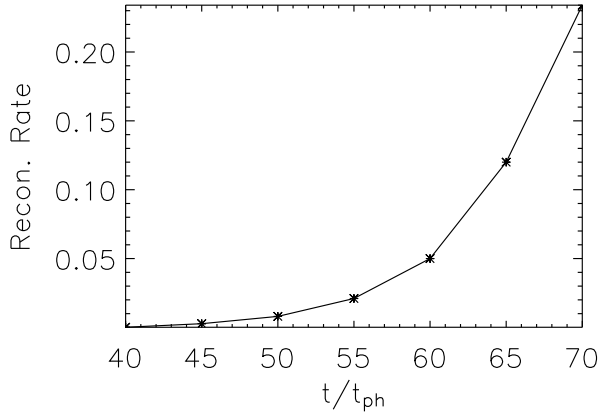}}
 \caption{As the arcade of the emerging tube expands into the magnetized corona, the continued weakening of the overlying restraining field results in faster reconnection.}
\label{recon}
\end{figure}

\begin{figure}
 \resizebox{\hsize}{!}{\includegraphics[scale=0.5]{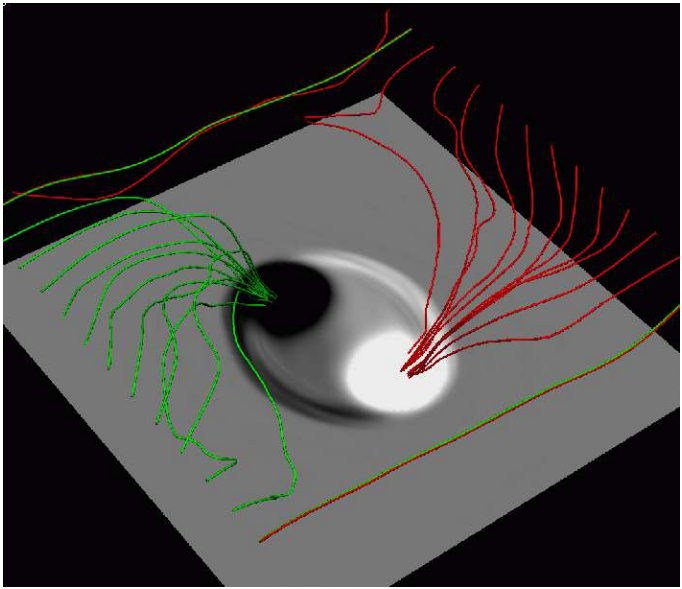}}
 \caption{The field line structure of the corona after external reconnection.  Red and green coronal lines are traced from opposite sides of the box, $x=-50$ and $x=50$, respectively,  at $z=22$ and $t=70$.  A magnetogram shows $B_z$ at $z=0$.}
\label{corona}
\end{figure}

As mentioned previously, the pre-existing field of the corona is almost antiparallel to the field of the emerging tube.  As the tube emerges, the arcade it forms first makes contact with the coronal field at $t\approx 33$.  It pushes into the horizontal field and an arched current sheet forms between them.  Reconnection occurs and the outer field lines of the arcade, that pass through the current sheet, change their connectivity and connect to the coronal field.  The form of reconnection is fully 3D in the sense that it does not involve a null point.  \citet{maclean09} also find this to be the case, with two clusters of null points forming at the sides of the emerging arcade rather than at the apex.  They suggest that separator reconnection occurs in the current sheet between the tube and the corona.  The effect, however, of this \emph{external} reconnection is to weaken the tension of the coronal field.  This becomes increasingly rapid with the loss of the restraining field, as in the breakout model \citep{lynch08}.  Evidence for this can be found by looking at the reconnection rate as it varies in time.  The rate of reconnected flux is given by
\[
 \frac{{\rm d}\Phi_{\rm rec}}{{\rm d}t} = \int{E_\parallel}\,{\rm d}l,
\]
where the right hand side is the integrated parallel electric field along the reconnection line \citep{schindler08}.  In 3D, however, no unique line exists at which the flux is split and reconnected \citep{hornig03}.  Within the current sheet, every field line constantly changes its connection.  Therefore, to estimate the reconnection rate between the flux tube arcade and the corona, in the simulation, we measure
\[
\frac{{\rm d}\Phi_{\rm rec}}{{\rm d}t} \approx \max_{y=0}(E_\parallel)\delta,
\]
where $\max_{y=0}(E_\parallel)$ is the maximum parallel electric field in the $y=0$ plane (at the top of the arched current sheet) and $\delta$ is the thickness of the current sheet.  This is a conservative estimate of the reconnection rate, giving a lower bound.  Figure \ref{recon} displays the increasing external reconnection rate during the expansion of the emerging arcade.  \citet{lynch08} report that this behaviour is also found in the breakout model.  Figure \ref{corona} shows the field line structure of the reconnected corona at $t=70$.  The coronal field lines are traced from opposite sides of the computational box (red for one side $(x=-50)$, green for the other $(x=50)$) at $z=22$.  A magnetogram showing $B_z$ is placed at the base of the photosphere $(z=0)$.  After external reconnection the coronal field connects down into the main photospheric polarities (sunspots), leaving the centre free for the arcade to push on upwards.

As the arcade emerges, the Lorentz force produces shearing along the polarity inversion line between the two sunspots.  Also during the rise of the arcade, plasma drains down the field lines, with some flowing into a region of reduced pressure within the arcade.  This, combined with the shearing produced by the Lorentz force, results in \emph{internal} reconnection in the arcade and the production of a flux rope.  In the simulation this occurs at $t\approx 80$.  For more details of the physics of the internal reconnection and flux rope production, the reader is directed to \citet{dmac09b} \citep[see also][]{manchester04}.

With the production of a flux rope and the continued loss of the restraining coronal field through external reconnection, there is a catastrophic expulsion of the flux rope at $t\approx 91$.  This description of the eruption is very similar to that of magnetic breakout.  There are two important differences, however.  The first is that in this flux emergence model, a null point does not play a crucial role in the external reconnection, as it does in breakout.  The second is that there are no imposed flows in the flux emergence model.  The Lorentz force of the emerging arcade naturally shears the magnetic field \citep{dmac09b}.

\begin{figure}
 \resizebox{\hsize}{!}{\includegraphics{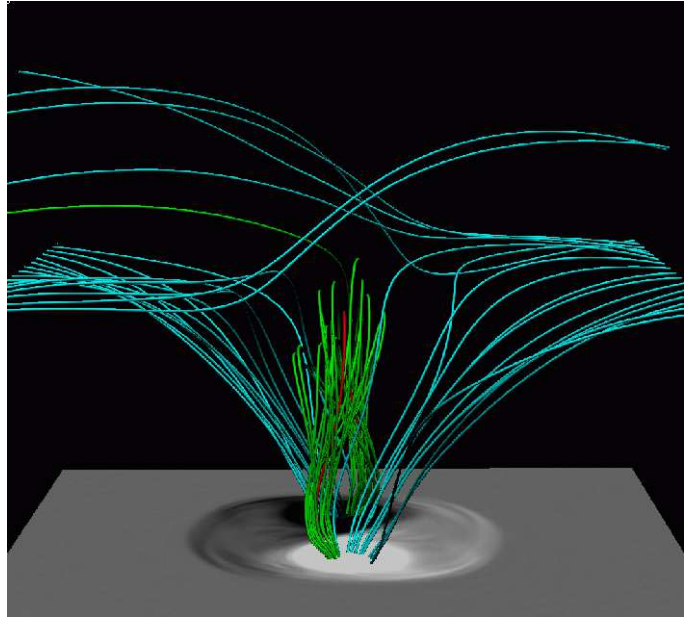}}
 \caption{The field line configuration just before the eruption of the second rope.  Coronal field lines (cyan) are traced from the sides of the box (along the $y$-axis) at $z=40$.  The red field line is the axis of the newly formed flux rope.  Some field lines surrounding this are traced in green.  A magnetogram is displayed at $z=0$.}
\label{takeoff}
\end{figure}

After the first eruption, the system does not settle into an equilibrium since the flux tube continues to emerge. The magnetic field in the corona is no longer horizontal due to reconnection during the first eruption.  Shearing in the emerging arcade produces a second flux rope at a height $z\approx 34$ and time $t\approx 125$.  The field line structure of the new flux rope and the corona at $t=125$ is displayed in Figure \ref{takeoff}.  Coronal field lines (cyan) are traced from opposite sides of the box at $z=40$.  The red field line is the axis of the new flux rope.  Some surrounding field lines (green) are traced from the main photospheric polarities, which are shown on a magnetogram at $z=0$.  This flux rope also erupts but the mechanism of the eruption is different to that of the first.  The first eruption was of breakout type where significant external reconnection (between the emerging arcade and the horizontal coronal field) played a crucial role in the expansion phase up to the eruption.  As can be seen in Figure \ref{takeoff}, however, the first eruption has `cleared a path' in the corona.  External reconnection, therefore, does not feature as a key factor in the eruption of the second rope.  This means that the trigger mechanism is not of breakout type and other candidates need to be considered. \citet{archontis08} suggest that the eruption they found is mainly driven by runaway reconnection below the flux rope.  This process has two important effects on the eruption.  Firstly, the reconnection weakens the tension of the overlying field.  Secondly, the upward reconnection jet carries the reconnected field lines to the erupting flux rope, adding poloidal flux to the rope \citep{vrsnak08}.  Both of these effects help the rope to accelerate further.  This, in turn, enhances the reconnection below and a runaway process ensues.  For the second flux rope, this in combination with a weakened coronal field could drive the eruption.  Another possibility, however, is that the newly formed rope becomes subject to an ideal MHD instability.  A likely candidate is the \emph{torus instability} \citep{bateman78,kliem06,torok07}. A toroidal flux tube will become unstable against expansion if the external poloidal field decreases sufficiently rapidly in the direction of the major tube radius. \citet{fan09} simulates the emergence of a cylindrical flux tube into a non-magnetized corona.  The author claims that the expansion and acceleration of the flux rope, produced in the simulation, is due to the continuous injection of twist, from the interior, via torsional Alfv\'{e}n waves.

Identifying the trigger mechanism for the second eruption is not a trivial undertaking and is therefore beyond the scope of this research note. This is a subject for future study.  It is possible that there is no single trigger mechanism and that the second eruption is due to a combination of mechanisms, including those mentioned above.

\begin{figure}
 \resizebox{\hsize}{!}{\includegraphics{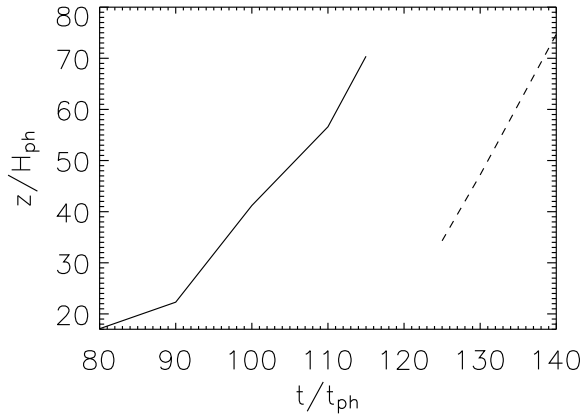}}
 \caption{The height-time profiles for the two eruptions of the simulation.  The curve for the first eruption is solid and the second is dashed.}
\label{ht}
\end{figure}

\begin{figure}
 \resizebox{\hsize}{!}{\includegraphics{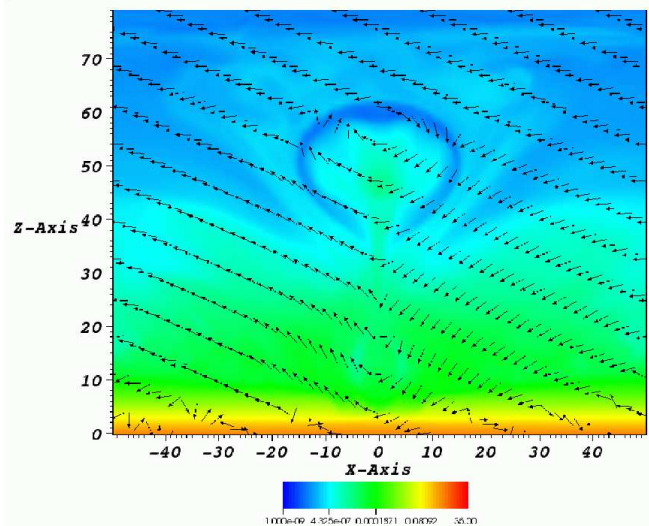}}
 \caption{The second flux tube eruption of the simulation at $t=132$.  The colour map shows $\log\rho$ in the $y=0$ plane with field line arrows to show the direction (but not the magnitude) of the magnetic field.  Dense plasma can clearly be seen to be carried upwards into the corona with the erupting flux tube.}
\label{eruption}
\end{figure}

The height-time profiles of the erupting ropes are displayed in Figure \ref{ht}.  These are estimated by tracking the O-point of the magnetic field in the $y=0$ plane.  The O-points also carry a `circular' profile of dense plasma so we are confident that these accurately represent the erupting flux ropes.  This is demonstrated in Figure \ref{eruption}, which shows the second erupting flux rope, with centre $(x,z)\approx(0,55)$, at $t=132$ by displaying the plasma density in the $y=0$ plane.  Field line arrows are also shown to indicate the direction of the magnetic field in that plane.

As mentioned earlier, the first flux rope forms at $t\approx 80$.  It rises slowly until the eruption at $t\approx 91$.  Here the gradient of the curve, in Figure \ref{ht}, changes from (in non-dimensionalized units) 0.3 to 2.  At $t=115$ the centre of the rope has reached $z=70$ and the gradient of the curve has increased to 2.8.  We stop tracking the rope at this time since this is just before it comes into contact with the top boundary of the computational box.  At $t\approx125$ the second rope, that has formed, is ejected upwards.  Again, the rope is only tracked to just above $z=70$ since beyond this the upper boundary begins to interfere with the rope's ascent.

The simulation is ended after the second eruption due to computational constraints.  The plasma from the first two eruptions hits the top of the computational box, falls back down and interferes with the system.  In theory, however, as long as the emergence process transports enough flux into the atmosphere and drives the necessary shearing, there should be more eruptions like the ones described.  

\begin{figure}
 \resizebox{\hsize}{!}{\includegraphics{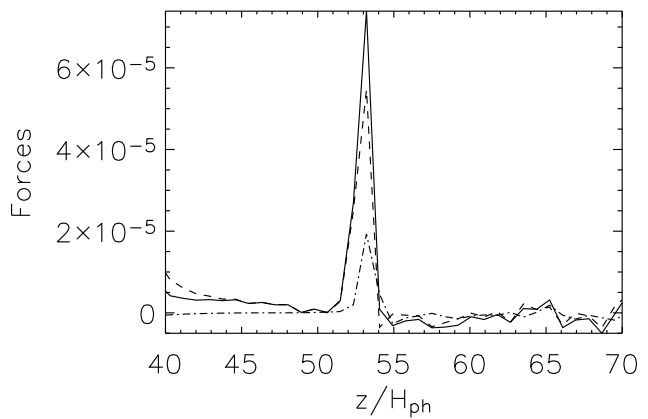}}
 \caption{The vertical forces carrying the flux rope upwards at $t=130$.  The cut is taken at $(x,y)=(0,0)$.  Key: All forces (solid), $(\mathbf{j}\times\mathbf{B})_z$ (dash), $-\partial p/\partial z$ (dot-dash).}
\label{forces}
\end{figure}

The ropes are carried upwards by a Lorentz force.  Looking at the vertical forces,  $(\mathbf{j}\times\mathbf{B})_z - \partial p/\partial z - \rho g > 0$ at the height of a flux tube during its rise.  Figure \ref{forces} illustrates this by displaying the (non-dimensionalized) vertical forces as a function of height at $(x,y) = (0,0)$ for the second flux rope eruption at $t=130$.  A positive upward force exists at the height of the flux rope and moves upwards in time with it.  Of all the forces, it is the Lorentz force that dominates and is ultimately responsible for the rise of the ropes.

\section{Summary}
In this paper we have demonstrated that multiple CME-like eruptions are possible from a toroidal flux tube emerging into a magnetized corona.  This combines and builds on the work of \citet{dmac09b} and \citet{archontis08}.  For the present study we consider a corona with a field that is almost antiparallel to the field of the emerging tube.  External reconnection at the apex of the emerging arcade weakens the tension of the coronal field.  With the expansion of the arcade, this reconnection becomes faster through time. Shearing, which occurs as part of the emergence process, induces internal reconnection in the arcade and produces a flux rope.  The continued emergence in combination with removal of the overlying coronal field eventually results in the expulsion of the flux rope. The mechanism for this eruption is similar to that of the breakout model.  One important difference, however, is that the external reconnection in this model does not take place at a null point.
After the first eruption, continued emergence and, therefore, shearing results in the formation of a second flux rope.  This also erupts but the trigger mechanism cannot be directly linked to the breakout model, as with the first eruption.  Due to the reconnection of the first eruption with the corona, a weakened coronal field exists above the second rope when it forms.  Possible trigger mechanisms, such as runaway reconnection and the torus instability, have been suggested.  However, it is possible that the trigger for the second eruption is a combination of such mechanisms.

\begin{acknowledgements}
DM and AWH would like to thank the referee, Tibor T{\"o}r{\"o}k, whose comments improved this research note. DM acknowledges financial assistance from STFC.  The computational work for this paper was carried out on the joint STFC and SFC (SRIF) funded cluster at the University of St Andrews.  DM and AWH acknowledge financial support from the European Commission through the SOLAIRE Network (MTRN-CT-2006-035484).
\end{acknowledgements}

\end{document}